\documentstyle[aps,preprint,epsfig]{revtex}

\begin{document}

\draft

\title{
High-sensitivity force measurement using entangled probes}

\author{
Stefano Mancini and Paolo Tombesi
}

\address{
INFM, Dipartimento di Fisica,
Universit\`a di Camerino,
I-62032 Camerino, Italy}

\date{\today}

\maketitle

\widetext

\begin{abstract}
We show the possibility to improve the 
measurement sensitivity of a weak force
by using massive probes in an entangled state. 
This latter can be achieved by exploiting radiation 
pressure effects. 
\end{abstract}


Since the original formulation of quantum mechanics,
entanglement has been recognized as one of its most 
puzzling features 
\cite{SCH,EPR}.
During the last few years a considerable amount of literature focused 
on 
methods to prepare atoms in nonclassical and
also entangled states appeared
\cite{ATOMS}. A striking achievement in this rapidly 
expanding field has been the recent entanglement
of two gas samples of atoms \cite{BP}.
Contemporarily we  
proposed to exploit the
radiation pressure to entangle massive macroscopic oscillators, like
pendular mirrors in a cavity \cite{MACRO}.
Once the entanglement between macroscopic oscillators
has been shown to be accessible, another important achievement would be 
to give a technological application of 
such a property of their states, by considering them as probes. Then, 
our aim is to show here the 
potentiality of using these entangled oscillators in quantum metrology. 
This was already 
recognized in Ref. \cite{WIN}, where 
it was
proposed to use entangled atoms for very high precision clocks.
We shall show, instead, that one can use entangled probes for 
very sensitive tiny forces measurements.

Essentially, a classical coherent force causes a displacement
of the probe (e.g. a pendular mirror)
resulting in the shift of its state in the phase space.
Such a displacement would be recognized through a 
readout apparatus (e.g. radiation field) \cite{BRAG}.
Therefore, the ultimate limit in this kind of measurement
is represented by the width of the probe's probability density
along the displacement direction. In the most performing case, 
the width of the oscillator's ground state
probability density determines the so called
Standard Quantum Limit (SQL) \cite{BRAG}.
This state can be identified with the vacuum state
at zero temperature, so, to beat it, 
one should put the probe at least in a squeezed state \cite{HOLL}.

Cavities with one pendular mirror have already been studied 
\cite{VAR}, showing a variety of intriguing effects
coming from optomechanical coupling \cite{NCS}. 
Moreover, due to the recent technological developments 
in optomechanics, this area is now
experimentally accessible \cite{EXP}.
Therefore, the aim of this paper is to show that the use of two 
entangled probes (two pendular mirrors) 
surely improves the sensitivity of a weak
force measurement.

\vspace{1cm}

{\it State preparation}

\vspace{0.5cm}

Let us consider a cavity with two (identical) pendular mirrors, 
playing the role of probes,
whose free Hamiltonian can be written as 
\begin{equation}\label{eq:H0}
    {\hat H}_{0}=\hbar\Omega\sum_{j=1}^{2}\left
    (\frac{{\hat p}_{j}^{2}}{2}+\frac{{\hat q}_{j}^{2}}{2}\right)\,,
\end{equation}
where ${\hat q}_{j}$
and ${\hat p}_{j}$ ($j=1,2$)
are the (dimensionless)
position and momentum 
operators (with commutation relation
$[{\hat q}_{j},{\hat p}_{k}]=i\delta_{jk}$) 
and $\Omega$ is their frequency. 

In practical situations massive probes 
are always far from SQL \cite{BRAG}, 
mainly due to the thermal noise 
associated to their (initial) state.
Here, we shall show how to prepare 
a probes' state having a reduced uncertainty which 
improves their sensitivity.
To this end we recast the model introduced in 
Ref.\cite{MACRO}, where 
we have shown the possibility of 
entangling massive probes by exploiting 
the radiation pressure.
We are going to consider a radiation field,
the {\it entangler}, 
described by the cavity mode ${\hat b}$,  
mediating information between the two mirrors (probes). 
The interaction Hamiltonian, to be added to Eq.(\ref{eq:H0}), 
results of the form 
\begin{equation}\label{eq:Hint1}
    {\hat H}_{int}=\hbar G {\hat b}^{\dag}{\hat b}
    \left({\hat q}_{1}-{\hat q}_{2}\right)\,,
\end{equation}
where $G$ is the optomechanical coupling constant 
depending on the probes' mass \cite{VAR}.
Since we require an intense cavity field we are naturally lead
to a quadratic form of ${\hat H}_{int}$ by the simple replacement
${\hat b} \to \beta+{\hat b}$, 
where now $\beta$ denotes the classical amplitude 
of the cavity field and $\hat{b}$ the quantum fluctuation.
Then, we have the following linear Heisenberg equations
\begin{mathletters}\label{eqs:LINEQS}
\begin{eqnarray}
    \dot{\hat b} &=& i\Delta {\hat b}
    -i G\beta ({\hat q}_1-{\hat q}_2) \,,
    \\
    \dot{\hat q}_j &=& \Omega {\hat p}_j \,,
    \\
    \dot{\hat p}_j &=& -\Omega {\hat q}_j
    +(-)^j 2 G (\beta^* {\hat b}+\beta {\hat b}^{\dag})
    \,,
\end{eqnarray}
\end{mathletters}
where $j=1,2$ and $\Delta$
is the field detuning \cite{MACRO}. 
For a sufficiently large value of $\Delta$ the ${\hat b}$ mode 
can be adiabatically eliminated obtaining
\begin{mathletters}\label{eqs:EQSMET}
\begin{eqnarray}\label{eqs:qp}
    \dot{\hat q}_j &=& \Omega {\hat p}_j \,,
    \\
    \dot{\hat p}_j &=& -\Omega {\hat q}_j
    +(-)^j \frac{(2G|\beta|)^{2}}{\Delta}
    \left({\hat q}_{1}-{\hat q}_{2}\right) \,.
\end{eqnarray}
\end{mathletters}
The solution of these equations reads
\begin{eqnarray}\label{eqs:SOL}
    {\hat q}_{j}(t)&=&\frac{1}{2}\left[
    \cos(\Omega t)+(-)^{j-1}\cos(\Theta t)\right] {\hat q}_{1}(0)
    +\frac{1}{2}\left[
    \sin(\Omega t)+(-)^{j-1}
    \frac{\Omega}{\Theta}\sin(\Theta t)\right] {\hat p}_{1}(0)
    \nonumber\\
    &&+\frac{1}{2}\left[
    \cos(\Omega t)+(-)^{j}\cos(\Theta t)\right] {\hat q}_{2}(0)
    +\frac{1}{2}\left[
    \sin(\Omega t)+(-)^{j}
    \frac{\Omega}{\Theta}\sin(\Theta t)\right] {\hat p}_{2}(0)
    \,,
\end{eqnarray}
where we have introduced the quantity 
\begin{equation}\label{eq:The}
    \Theta=\left\{\Omega\left[\Omega+2\frac{(2G|\beta|)^{2}}{\Delta}
    \right]\right\}^{1/2}\,.
\end{equation}
The corresponding expression for ${\hat p}_{j}(t)$ can be easily determined
through the time derivative of the above Eq.(\ref{eqs:SOL}).

Now, we evaluate the fluctuations 
over a separable thermal state 
as could be that of realistic massive probes.
Such state can be written as 
\begin{equation}\label{eq:rhoth}
    {\hat\rho}_{th}=
    {\cal Z}\exp\left[-\frac{\hat{H}_{0}}{k_{B}T}\right]\,,
\end{equation}
where ${\cal Z}=\left[
\sum_{n}\exp(-n\hbar\Omega/k_{B}T)\right]^{-2}$ with $k_{B}$ the 
Boltzmann constant and $T$ the equilibrium temperature.
The average number of thermal excitations for each probe
is ${\cal N}_{th}=[\coth(\hbar\Omega/2k_{B}T)+1]/2$ \cite{VIT}.
Then, it can be easily seen that
if we turn off the entangler at time $t=\pi/(2\Theta)$, 
we are left with the following variances
\begin{equation}\label{eq:corr}
    {\bf C}\equiv
    \frac{1}{2}\langle (\hat{\bf v}\,\hat{\bf v}^{\,T})
    +(\hat{\bf v}\,\hat{\bf v}^{\,T})^{T}
    \rangle_{{\hat\rho}_{th}}
    =\frac{1}{2}\left(\frac{1}{2}+{\cal N}_{th}\right)
    \left(\begin{array}{cccc}
    1+r^{-2}&0&1-r^{-2}&0
    \\
    0&1+r^{2}&0&1-r^{2}
    \\
    1-r^{-2}&0&1+r^{-2}&0
    \\
    0&1-r^{2}&0&1+r^{2}
    \end{array}
    \right)\,,
\end{equation}
where ${\hat{\bf v}}^{\,T}=({\hat q}_{1},{\hat p}_{1},
{\hat q}_{2},{\hat p}_{2})$
and $r=\Theta/\Omega\ge 1$.
By using Eq.(\ref{eq:corr}) we recognize that for $r \gg 1$ it 
is possible to reduce the noise in the quadrature 
${\hat q}_{1}-{\hat q}_{2}$ 
bringing the probes state close to an eigenstate
of this quadrature. 
In such a case the  
probes' state would be a kind of EPR state \cite{EPR}.
To be more precise,  
a general separability criterion introduced in Ref. \cite{MACRO} gives
\begin{equation}\label{eq:ent}
    \langle({\hat q}_{1}-{\hat q}_{2})^{2}\rangle_{{\hat\rho}_{th}}
    \times \langle({\hat p}_{1}+{\hat p}_{2})^{2}\rangle_{{\hat\rho}_{th}}
    \ge |\langle[{\hat q}_{j},{\hat p}_{j}]\rangle_{{\hat\rho}_{th}}|^{2}\,,
\end{equation} 
then, the condition
$r^{2}>(1+2{\cal N}_{th})$ 
guarantees entanglement.
This leads to a great improvement in the minimum detectable force
as we shall see.

Furthermore, after interaction (\ref{eq:Hint1}), 
leaving the probes to freely evolve  
one ends up with a Gaussian state, say $\hat{\rho}_{12}$,
characterized by the following covariance matrix 
\begin{equation}\label{eq:corrphi}
    {\bf C}(\phi)={\bf R}(\phi){\bf C}{\bf R}^T(\phi)\,,
    \quad
    {\bf R}(\phi)=\left(
    \begin{array}{cccc}
    \cos\phi&\sin\phi&0&0
    \\
    -\sin\phi&\cos\phi&0&0
    \\
    0&0&\cos\phi&\sin\phi
    \\
    0&0&-\sin\phi&\cos\phi
    \end{array}\right)\,,
\end{equation}
where $\phi$ represents the 
phase space rotation angle
depending on the frequency $\Omega$
and on the elapsed time.

\vspace{1cm}

{\it Weak force measurement}

\vspace{0.5cm}

After state preparation,
we consider the action of a classical force on the
probes and its 
readout through radiation fields.
Thus, the interaction 
Hamiltonian to be added to $H_{0}$ would now be
\begin{equation}\label{eq:Hint2}
    H_{int}=-\hbar\Omega f({\hat q}_{1}-{\hat q}_{2})
    -\hbar g \hat{c}_{1}^{\dag}c_{1}\hat{q}_{1}
    +\hbar g \hat{c}_{2}^{\dag}c_{2}\hat{q}_{2}\,,
\end{equation}
where $f$ is the dimensionless
force strength. Here, the force is considered as
acting on the relative variables of the probes,
as for example, 
it is the case of gravitational radiation \cite{GRAV}.
Furthermore, $\hat{c}_{j}$ are operators describing the 
meter fields which deserve as readout. These could be
proper modes of additional cavities.
They are optomechanically 
coupled to the probes with a strength $g$.

Assuming again intense fields, we are led to 
linear Heisenberg equations
\begin{mathletters}\label{eqs:lineqs}
\begin{eqnarray}
    \dot{\hat{c}}_{j}&=&i\Delta_{j}\hat{c}_{j}-(-)^j ig\gamma q_{j}\,,
    \\
    \dot{\hat{q}}_{j}&=&\Omega \hat{p}_{j}\,,
    \\
    \dot{\hat{p}}_{j}&=&-\Omega \hat{q}_{j}
    -(-)^j 2g \left(\gamma^{*}\hat{c}_{j}+\gamma\hat{c}_{j}^{\dag}\right)
    -(-)^j \Omega f\,,
\end{eqnarray}
\end{mathletters}
where $\gamma$ denotes the classical fields amplitude 
and $\hat{c}_{j}$ the quantum fluctuations.
Tpically, it should be $|\gamma|\ll|\beta|$ since 
radiation pressure could blur the signal.
That is also the reason why we now consider the entangler turned off.

Choosing $\Delta_{j}=0$, $\gamma \in {\bf R}$, and introducing the 
field quadratures 
${\hat X}_{j}=(\hat{c}_{j}+\hat{c}_{j}^{\dag})/\sqrt{2}$,
${\hat Y}_{j}=-i(\hat{c}_{j}-\hat{c}_{j}^{\dag})/\sqrt{2}$,
(with commutation relation $[\hat{X}_{j},\hat{Y}_{k}]=i\delta_{jk}$),
we immediately see that only the phase quadratures $\hat{Y}_{j}$
carrie out information about the force.
As matter of fact, from Eqs.(\ref{eqs:lineqs}), we obtain
\begin{eqnarray}\label{eqs:sol}
    \hat{Y}_{j}(\tau)&=&-(-)^j
    \frac{g\gamma}{\Omega}\sin\left(\Omega\tau\right)
    \hat{q}_{j}(0)
    -(-)^j\frac{g\gamma}{\Omega}
    \left[1-\cos\left(\Omega\tau\right)\right]
    \hat{p}_{j}(0)
    \nonumber\\
    &+&2\frac{(g\gamma)^{2}}{\Omega^{2}}\left[\Omega\tau
    -\sin\left(\Omega\tau\right)\right]
    \hat{X}_{j}(0)
    +\sqrt{2}\frac{g\gamma}{\Omega}\left[\Omega\tau-
    \sin\left(\Omega\tau\right)\right]f
    +\hat{Y}_{j}(0)\,,
\end{eqnarray}
where $\tau$ is the time duration of the force.
Then, performing homodyne detection \cite{hom} on the meters fields
one can get the following signal
\begin{equation}\label{eq:sig}
    \langle\hat{Y}_{1}(\tau)+\hat{Y}_{2}(\tau)\rangle
    \equiv {\cal S}(\tau)\, f
    =2\sqrt{2}\frac{g\gamma}{\Omega}\left[
    1+\Omega\tau-\cos\left(\Omega\tau\right)\right]\,f\,.
\end{equation}
The corresponding noise can be calculated by means of 
the initial state of the probes, $\hat{\rho}_{12}$,
and the vacuum noise for the meters.
Thus, we get
\begin{eqnarray}\label{eq:noi}
    \left\langle\left[\hat{Y}_{1}(\tau)+\hat{Y}_{2}(\tau)\right]^{2}
    \right\rangle
    &\equiv& {\cal N}(\tau)
    \nonumber\\
    &=&\left(\frac{g\gamma}{\Omega}\right)^{2}
    \sin^{2}\left(\Omega\tau\right) 
    \left[r^{-2}\cos^{2}\phi+r^{2}\sin^{2}\phi\right]
    \left(1+2{\cal N}_{th}\right)
    \nonumber\\
    &+&\left(\frac{g\gamma}{\Omega}\right)^{2}
    \left[1-\cos\left(\Omega\tau\right)\right]^{2}
    \left[r^{-2}\sin^{2}\phi+r^{2}\cos^{2}\phi\right]
    \left(1+2{\cal N}_{th}\right)
    \nonumber\\
    &-&2\left(\frac{g\gamma}{\Omega}\right)^{2}\sin\left(\Omega\tau\right)
    \left[1-\cos\left(\Omega\tau\right)\right]
    \left[r^{-2}-r^{2}\right]\sin\phi\cos\phi
    \left(1+2{\cal N}_{th}\right)
    \nonumber\\
    &+&4\left(\frac{g\gamma}{\Omega}\right)^{4}
    \left[\Omega\tau-\sin\left(\Omega\tau\right)\right]^{2}
    \nonumber\\
    &+&1\,.
\end{eqnarray}
The first three terms in Eq.(\ref{eq:noi}) 
come from the probes state, while the fourth term is 
due to the back action of the meters modes (radiation pressure noise),
and the final $1$ is the shot noise term. 
It is worth noting that for $\Omega\tau$ integer multiple
of $2\pi$ the noise does no longer depend on the initial probes
state \cite{BRAGETAL}, 
but for all other times it does. 
Such state in our case is characterized by ${\cal N}_{th}$, $r$
and $\phi$. Assuming ${\cal N}_{th}$ fixed, one might 
optimize $\phi$, and then reduce the noise by increasing
$r$. In doing so we get
\begin{equation}\label{eq:phiopt}
    \phi_{opt}=-\frac{1}{2}\arctan\left\{
    -2\frac{\sin(\Omega\tau)\left[1-\cos(\Omega\tau)\right]}
    {\sin^{2}(\Omega\tau)-\left[1-\cos(\Omega\tau)\right]^{2}}
    \right\}+n\frac{\pi}{2}\,,
\end{equation} 
where $n$ is an integer number.

The relevant quantity to study 
is the signal to noise ratio
\begin{equation}\label{eq:SNR}
    {\cal R}=\frac{|{\cal S}|}{\sqrt{\cal N}}f\ge 1\,,
    \quad\Longrightarrow\quad
    f_{min}=\frac{\sqrt{\cal N}}{|{\cal S}|}\,,
\end{equation}
which gives the minimum detectable force.
The latter, for ${\cal N}_{th}=0$ and $r=1$,
represents the SQL.

\begin{figure}[h]
\vspace{0.5cm}
\centerline{\epsfig{figure=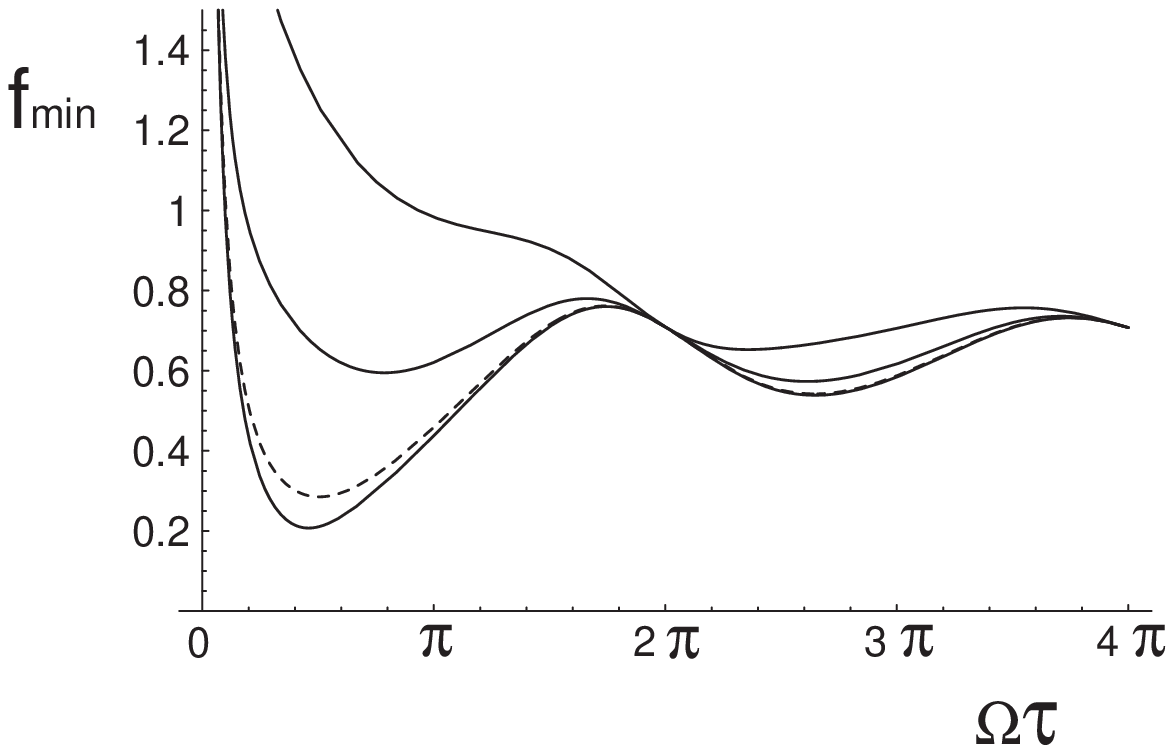,width=3.5in}}
\vspace{0.5cm}
\caption{\widetext 
The minimum detectable force $f_{min}$ is plotted versus 
the scaled time $\Omega\tau$ for ${\cal N}_{th}=20$
and $(g\gamma/\Omega)=1$.
Solid curves from top to bottom are for 
$r=1$, $r=2$ and $r=10$. 
Instead, the dashed curve represents the SQL.
}
\label{fig1}
\end{figure}

\begin{figure}[h]
\vspace{0.5cm}
\centerline{\epsfig{figure=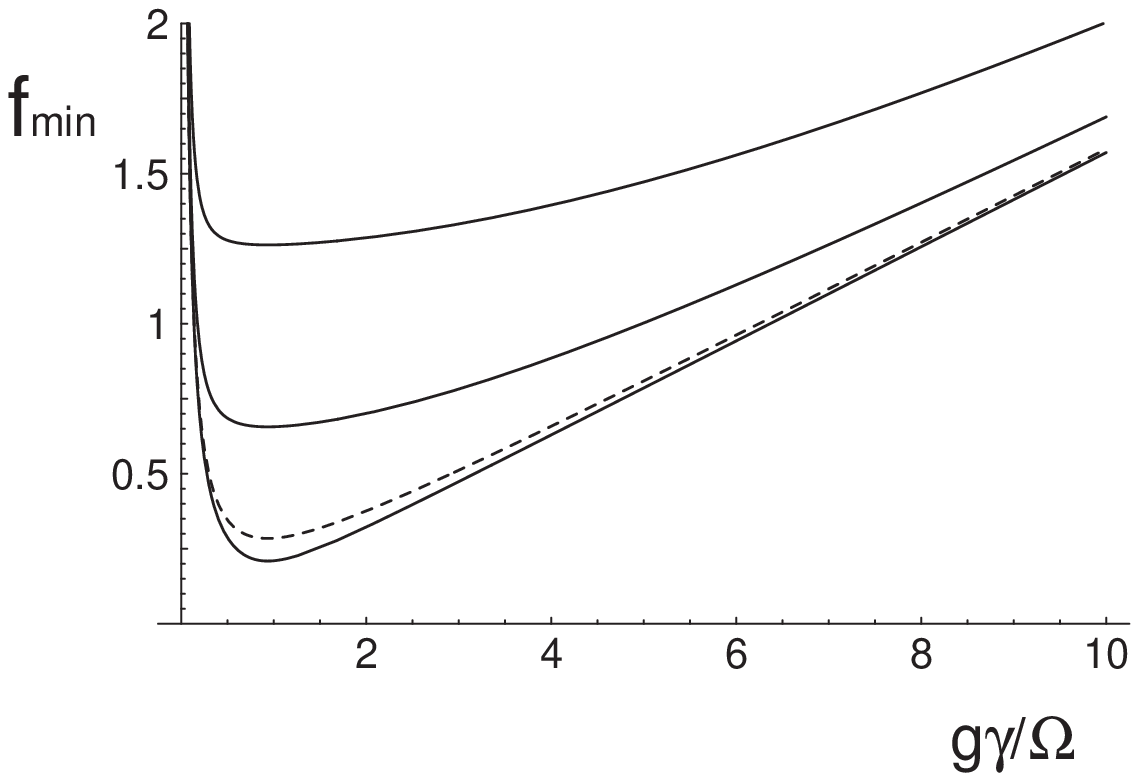,width=3.5in}}
\vspace{0.5cm}
\caption{\widetext 
The minimum detectable force $f_{min}$ is plotted versus 
the quantity $g\gamma/\Omega$ for ${\cal N}_{th}=20$
and $\Omega\tau=\pi/2$.
Solid curves from top to bottom are for 
$r=1$, $r=2$ and $r=10$. 
Instead, the dashed curve represents the SQL.
}
\label{fig2}
\end{figure}

In Fig.\ref{fig1} we show $f_{min}$ as function of 
the (scaled) time when the force is on.
The improved sensitivity due to the probes correlations 
is especially evident for $\Omega\tau < 2\pi$.
Above this limit, the meters back action noise 
becomes dominant, thus reducing the benefit.
Such back-action noise could be reduced by 
decreasing the value of $(g\gamma/\Omega)$,
i.e., the intensity of the meter fields, 
but this also affects the strength of the signal.
Thus, it turns out that, for each $\tau$, there exist an
optimum value of $(g\gamma/\Omega)$
as shown in Fig.\ref{fig2}.
Also notice in Figs.\ref{fig1} and \ref{fig2} 
that the entanglement condition, 
$r^{2}>(1+2{\cal N}_{th})$,
alllows to beat the SQL (bottom curves).
With the present model, the sensitivity of the force 
measurement crucially depends on the value of $r$.
The latter is related the optomechanical coupling constant $G$ 
and on the intensity of the entangler field $|\beta|^{2}$. 
For instance, the parameters values used in
experiments of Ref.\cite{EXP} 
permit to reach $r \approx 50$,
which would allow to beat the SQL, 
even with ${\cal N}_{th}=10^{3}$.

It is also to remark that the achievable improvements 
are not uniform in the phase space, since we have assumed the 
phase $\phi$ locked to the time duration $\tau$ of the force 
in Eq.(\ref{eq:phiopt}).
Thus, we have a kind of two-mode squeezed state,
and the situation is analogous to the use of 
a single-mode squeezed state,
where the direction of the reduced uncertainty should 
coincide with the one of the force displacement \cite{HOLL}.
This enforces the conclusions of Ref.\cite{MUNRO}, 
about the equivalence of entanglement and squeeezing,
while the more optimistic results of Ref.\cite{MAURO}
are not applicable in this case.
We would remark however that 
probes' entangled state seems experimentally accessible 
with present technology while probes' squeezed state does not.

Finally, since the entangler 
has been considered turned off
prior the measurement,
what will really limit the applicability of 
the above procedure would be  
the decoherence, which degrades the 
prepared meters state.
Such a decoherence takes place in a time scale
$(\Gamma{\cal N}_{th})^{-1}$ \cite{VAR,hom}
with $\Gamma^{-1}$ the mechanical relaxation time
of the probes. 
Therefore, the time for rotation $\phi$, together with
the duration $\tau$ of the force,
should be less than $(\Gamma{\cal N}_{th})^{-1}$.
Thus, the model would be suitable for weak but
impulsive forces \cite{PRAR}. 
Favorable conditions for its implementation could be found 
in Micro Electro-Mechanical Systems \cite{MEMS,CLELAND},
or Atomic Force Microscopes \cite{AFM},
where large mechanical quality factors are achievable.

Summarizing, we have proposed high-sensitivity force
measurement by exploiting nonclassical features of 
the probes, like entanglement.
That would allow us to overcome the uncertainty related to 
the probes state and even to beat the SQL. 
Our scheme, even if oversimplified, may open new perspectives
in quantum metrology \cite{CLELAND}, as well as
in gravitational waves detection \cite{GRAV}.

\section*{Acknowledgments}
This work has been supported by the INFM under the PAIS project 
``Entanglement and Decoherence".
S. M. gratefully acknowledges financial support from 
Universit\`a di Camerino under the project 
``Giovani Ricercatori".

\begin{references}
\bibitem{BRAGETAL}
See also,
V. B. Braginsky, {\it et al.}, gr-qc/0109003,
for a discussion on the dependence of the noise 
from the initial probes state.
\end{references}
\end{document}